\documentclass[journal, twoside, final]{IEEEtran}
\usepackage{amssymb, amsmath, amsthm, amsfonts}
\usepackage{multirow}
\usepackage{balance}
\usepackage{siunitx}
\usepackage{enumitem}
\usepackage[caption=false, font=scriptsize, labelfont=rm, textfont=rm]{subfig}
\usepackage{bm}
\usepackage{algorithm}
\usepackage{algorithmic}
\usepackage{array, graphicx, cite}
\usepackage{longtable, threeparttable}
\usepackage{makecell}

\usepackage{xcolor, url}

\allowdisplaybreaks[4]

\begin{document}

\title{Hybrid CNN-Transformer Based Sparse Channel Prediction for High-Mobility OTFS Systems}
\author{Zhaowei Guan, Wenkun Wen, Peiran Wu, Chen Wang, and Minghua Xia
\thanks{Received 29 September 2025; accepted 17 October 2025. This work was supported in part by the National Natural Science Foundation of China under Grant 62171486, and in part by the Guangdong Basic and Applied Basic Research Project under Grant 2022A1515140166. The associate editor coordinating the review of this article and approving it for publication was M. Derakhshani. \itshape{(Corresponding authors: Peiran Wu; Minghua Xia.)}}

%\thanks{Received 29 September 2025; accepted 17 October 2025. This work was supported in part by the Guangdong Basic and Applied Basic Research Project under Grant 2022A1515140166. The associate editor coordinating the review of this article and approving it for publication was M. Derakhshani. \itshape{(Corresponding authors: Peiran Wu; Minghua Xia.)}}

\thanks{Zhaowei Guan, Peiran Wu, and Minghua Xia are with the School of Electronics and Information Technology, Sun Yat-sen University, Guangzhou 510006, China (e-mail: guanzhw5@mail2.sysu.edu.cn, wupr3@mail.sysu.edu.cn, xiamingh@mail.sysu.edu.cn).} 
\thanks{Wenkun Wen is with the Techphant Technologies, Co. Ltd., Guangzhou 510310, China (email: wenwenkun@techphant.net)}
\thanks{Chen Wang is with the 54th Research Institute of China Electronics Technology Group Corporation, Shijiazhuang 050081, China (e-mail: wangc\_1994@163.com).} 	
}

\markboth{IEEE Wireless Communications Letters}{Guan \MakeLowercase{\textit{et al.}}: Hybrid CNN-Transformer Based Sparse Channel Prediction for High-Mobility OTFS Systems}

\maketitle

\IEEEpubid{\begin{minipage}{\textwidth} \ \\[12pt] \centering 2162-2345 \copyright\ 2025 IEEE. All rights reserved, including rights for text and data mining, and training of artificial intelligence \\ and similar technologies. Personal use is permitted, but republication/redistribution requires IEEE permission. \\
See \url{https://www.ieee.org/publications/rights/index.html} for more information.\end{minipage}}

\begin{abstract}
High-mobility scenarios in next-generation wireless networks, such as those involving vehicular communications, require ultra-reliable and low-latency communications (URLLC). However, rapidly time-varying channels pose significant challenges to traditional OFDM-based systems due to the Doppler effect and channel aging. Orthogonal time frequency space (OTFS) modulation offers resilience by representing channels in the quasi-static delay-Doppler (DD) domain. This letter proposes a novel channel prediction framework for OTFS systems using a hybrid convolutional neural network and transformer (CNN-Transformer) architecture. The CNN extracts compact features that exploit the DD-domain sparsity of the channel matrices, while the transformer models temporal dependencies with causal masking for consistency. Simulation experiments under extreme $500$ \si{km/h} mobility conditions demonstrate that the proposed method outperforms state-of-the-art baselines, reducing the root mean square error and mean absolute error by $12.2\%$ and $9.4\%$, respectively. These results demonstrate the effectiveness of DD-domain representations and the proposed model in accurately predicting channels in high-mobility scenarios, thereby supporting the stringent URLLC requirements in future wireless systems.
\end{abstract}


\IEEEpubidadjcol

\section{Introduction}

\IEEEPARstart{O}{rthogonal} time frequency space (OTFS) modulation has recently gained significant attention as a promising solution for reliable communications in high-mobility environments. By representing doubly selective (time- and frequency-selective) channels in the delay–Doppler (DD) domain, OTFS transforms rapid time–frequency variations into a quasi-static and sparse structure, making it well suited for vehicular, railway, and aeronautical scenarios where conventional orthogonal frequency division multiplexing (OFDM)-based methods often fail. To fully exploit this property, accurate DD-domain channel acquisition and prediction are indispensable.

A variety of channel estimation methods have been developed for OTFS systems. Model-driven neural estimators embed signal-processing priors into trainable architectures to improve robustness \cite{ModelDrivenOTFS}. Sparsity-aware deep methods leverage the inherent DD-domain sparsity for efficient recovery \cite{SparsePriorOTFS}, while tensor-based estimators enable scalable channel reconstruction in multiple-input multiple-output (MIMO) and millimeter-wave settings \cite{TensorOTFS}. In addition, intelligent reflecting surface (IRS)-assisted OTFS has been studied for enhancing high-mobility performance by jointly designing reflection patterns and channel estimators \cite{IRS-OTFS}. These works highlight the feasibility of channel acquisition in OTFS but primarily focus on one-shot estimation rather than predictive modeling.

In parallel, a large body of research has addressed wireless channel prediction using both statistical models and learning-based networks. Classical autoregressive (AR) processes, as well as deep architectures such as recurrent neural networks (RNNs), long short-term memory (LSTM) networks, multilayer perceptrons (MLPs), and Transformers, have been employed for time- or frequency-domain prediction \cite{CP-AR, CP-LSTM, CP-MLP, CP-Gruformer, CP-Overview}. While these methods demonstrate the benefits of data-driven prediction, they typically overlook the structural characteristics of OTFS and face scalability issues when applied to the high-dimensional DD-domain channel, which is represented as an $MN \times MN$ matrix.

Motivated by these limitations, this letter develops a compact, OTFS-aware channel prediction framework that explicitly exploits DD-domain sparsity and temporal correlation. The main contributions are summarized as follows:

\IEEEpubidadjcol

\begin{itemize}
	\item We design a hybrid convolutional neural network (CNN)–Transformer architecture tailored to high-mobility OTFS systems. Convolutional layers perform feature downsampling by exploiting DD-domain sparsity, Transformer layers capture long-range temporal dependencies with causal masking, and transposed convolutions reconstruct the predicted channel matrices.
	\item The framework mitigates the computational burden of high-dimensional DD-domain channels ($MN \times MN$) by reducing model complexity while preserving essential spatial and temporal structures.
	\item Simulation results under extreme $500$~\si{km/h} mobility confirm that the proposed lightweight DD-domain Transformer (LDformer) consistently outperforms strong baselines, achieving root mean square error (RMSE) and mean absolute error (MAE) reductions of $12.2\%$ and $9.4\%$, respectively, while supporting flexible multi-step prediction and scalable inference.
\end{itemize}
The remainder of the letter is organized as follows: Section II presents the OTFS system model and formulates the channel prediction problem. Section III elaborates on the proposed CNN-Transformer framework. Section IV discusses simulation results, followed by the conclusions in Section V.

{\it Notation:} Scalars, vectors, and matrices are denoted by regular letters (e.g., $m$), bold lowercase letters (e.g., $\bm{x}$), and bold uppercase letters (e.g., $\bm{H}$), respectively. The set $\mathbb{C}^{M \times N}$ denotes the space of complex-valued matrices of size $M \times N$. The superscript $(\cdot)^H$ indicates the Hermitian (conjugate transpose) operation. The operators $|\cdot|_F$ and $|\cdot|_1$ denote the Frobenius norm and $\ell_1$ norm, respectively. The function $\operatorname{vec}(\cdot)$ denotes vectorization, which stacks the columns of a matrix into a single column vector. The symbol $\otimes$ represents the Kronecker product. The matrix $\bm{F}_M$ denotes the $M \times M$ discrete Fourier transform (DFT) matrix, and $\bm{I}_M$ refers to the $M \times M$ identity matrix. Subscripts “DD”, “TF”, and “TD” refer to the delay-Doppler domain, time-frequency domain, and time domain, respectively.

\section{System Model}
\begin{figure}[!t]
\centering
\includegraphics[width=1.0\linewidth]{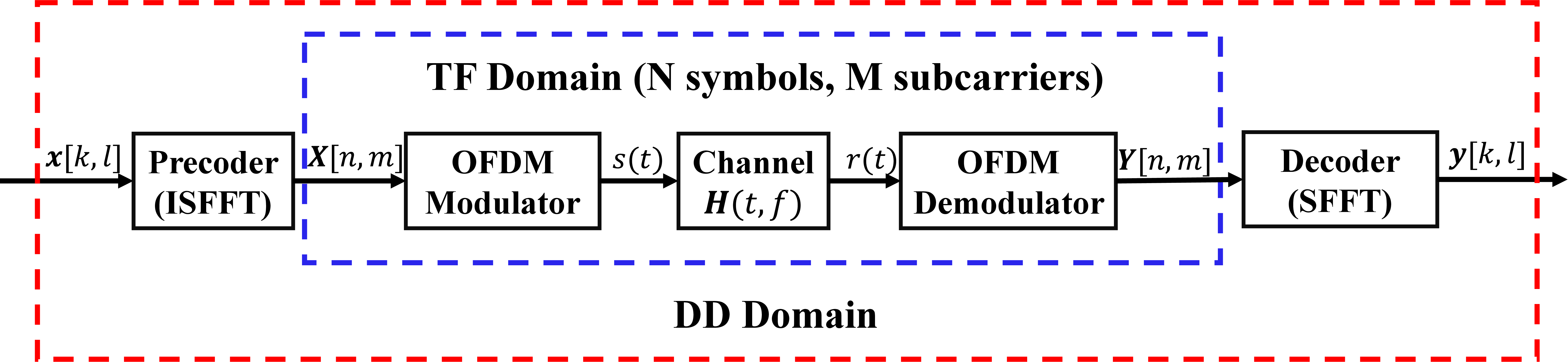}
\vspace{-15pt}
\caption{OTFS modulation system structure.}
\label{fig:OTFS_system}
\vspace{-10pt}
\end{figure}

Fig.~\ref{fig:OTFS_system} illustrates a high-mobility wireless system employing OTFS modulation. Information symbols are mapped from the DD domain to the time-frequency (TF) domain via the inverse symplectic finite Fourier transform (ISFFT), followed by OFDM modulation and transmission over time-frequency doubly selective channels. The received signal is demodulated and transformed back to the DD domain by the symplectic finite Fourier transform (SFFT) for detection \cite{OTFS}.

Let $M$ and $N$ denote the number of delay and Doppler bins, respectively. The DD-domain symbol matrix $\bm{X}{_\text{DD}} \in \mathbb{C}^{M \times N}$ is transformed to the TF domain via:
\begin{equation}
\bm{X}{_\text{TF}} = \bm{F}_M \bm{X}{_\text{DD}} \bm{F}_N^{H},
\end{equation}
where $\bm{F}_M$ and $\bm{F}_N$ are the DFT matrices of size $M \times M$ and $N \times N$, respectively. Subsequently, the time-domain transmit signal is given by:
\begin{equation}
\bm{s} = \operatorname{vec}(\bm{F}_M^{H} \bm{X}{_\text{TF}}) = (\bm{F}_N^{H} \otimes \bm{I}_M) \bm{x}_{\text{DD}},
\end{equation}
where $\bm{x}_{\text{DD}} = \operatorname{vec}(\bm{X}_{\text{DD}})$ is the vectorized DD-domain symbol.

After passing through a time-frequency doubly selective channel, the received signal in the time domain is:
\begin{equation}
\bm{r} = \bm{H}_{\text{TD}} \bm{s} + \bm{w},
\end{equation}
where $\bm{H}_{\text{TD}}$ is the time-domain channel matrix and $\bm{w}$ is a cyclically symmetric complex additive white Gaussian noise (AWGN).

Following demodulation and inverse symplectic transform, the received signal in the DD domain is:
\begin{equation}
\bm{y}_{\text{DD}} = (\bm{F}_N \otimes \bm{I}_M) \bm{r} = \bm{H}_{\text{DD}} \bm{x}_{\text{DD}} + \bm{w}_{\text{DD}},
\end{equation}
where $\bm{H}_{\text{DD}}$ is the effective DD-domain channel matrix, and $\bm{w}_{\text{DD}}$ is the transformed noise.

In high-mobility scenarios, channels vary rapidly over time, making the timely and accurate acquisition of CSI particularly challenging, especially for URLLC, where outdated CSI can severely degrade reliability and latency performance.

To address this issue, we formulate the {\it channel prediction problem} as follows: given $L$ historical DD-domain channel matrices ${\bm{H}_{t-L}, \cdots, \bm{H}_{t-1}}$ observed at time slots $t-L$ through $t-1$, the goal is to predict the channel at the current slot $t$:
\begin{equation}
\hat{\bm{H}}_t = f(\bm{H}_{t-L}, \bm{H}_{t-L+1}, \cdots, \bm{H}_{t-1}),
\end{equation}
where $\hat{\bm{H}}_t$ is the predicted channel and $f(\cdot)$ is the prediction function to be learned. Prediction accuracy directly impacts the system’s reliability, especially under fast channel variations.

While OTFS provides more stable channel representations than OFDM, it also introduces significant computational challenges. In OFDM systems, channel prediction typically involves $L \times M \times N$ matrices, where $M$ is the number of subcarriers and $N$ is the number of OFDM symbols. In contrast, the DD-domain channel in OTFS has size $MN \times MN$, leading to a prediction problem of scale $L \times MN \times MN$. This substantial increase in dimensionality renders conventional methods impractical, motivating the need for efficient learning-based solutions that can exploit the inherent structure of OTFS channel matrices.

\section{A Hybrid CNN-Transformer Architecture}
To address the challenge of channel prediction in high-mobility OTFS systems, we propose a hybrid CNN-Transformer architecture that exploits the latent, low-dimensional, and compact structure of DD-domain channel matrices. The design targets explicitly the computational burden introduced by the enlarged problem size in OTFS systems, leveraging the inherent sparsity nature of DD-domain channels to improve efficiency and prediction accuracy.

\begin{figure}[!t]
	\centering
	\includegraphics[width=0.8\linewidth]{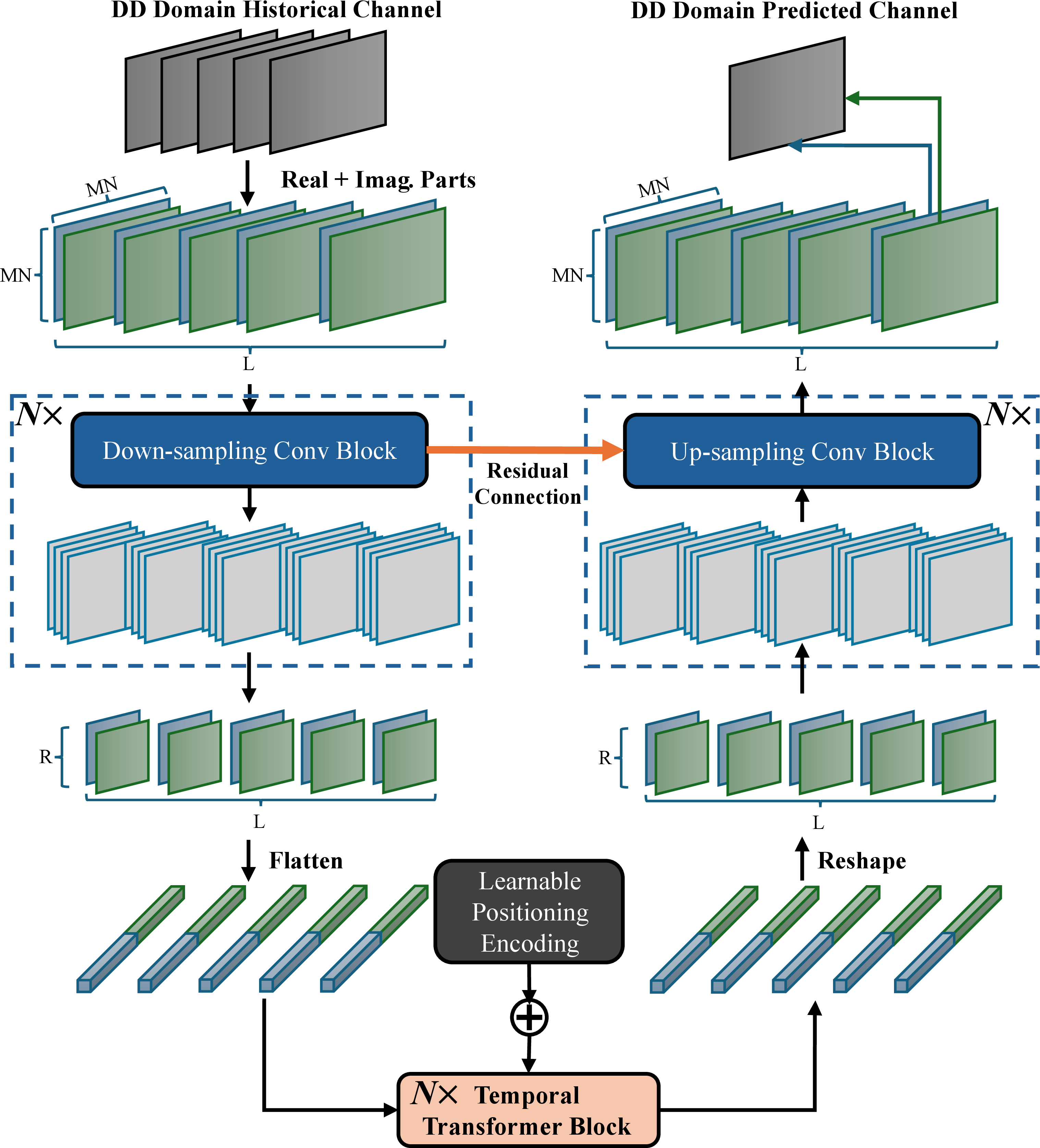}
	\vspace{-5pt}
	\caption{Hybrid CNN-Transformer architecture based on compact features that exploit DD-domain sparsity.}
	\label{fig:model_structure}
        \vspace{-10pt}
\end{figure}

As shown in Fig.~\ref{fig:model_structure}, the proposed framework consists of six sequential processing stages:
\begin{enumerate}[label = \arabic*)] 	
	\item Preprocessing: Historical DD-domain channel matrices are decomposed into their real and imaginary parts to form multi-channel input tensors.
	\item Compact Feature Extraction: Cascaded convolutional downsampling blocks compress the input and extract compact spatial features that exploit DD-domain sparsity.
	\item Temporal Modeling: The resulting feature sequences are passed through stacked Transformer encoder layers to capture temporal dependencies across time slots, with causal masking applied to preserve chronological order.
	\item Feature Map Reconstruction: Transformer outputs are reshaped to recover the 2D spatial layout of feature maps.
	\item Upsampling and Residual Learning: Transposed convolutional layers, equipped with residual connections, reconstruct the predicted DD-domain features.
	\item Prediction Output: Final channel matrices are generated by recombining the predicted real and imaginary components.
\end{enumerate}

Next, we begin with input preprocessing and data representation to elaborate on the key modules.

\subsection{Input Preprocessing and Data Representation}
The first step involves preprocessing the historical DD-domain channel matrices. Given $L$ complex-valued channel matrices ${\bm{H}_{t-L}, \bm{H}_{t-L+1}, \cdots, \bm{H}_{t-1}}$, we separate their real and imaginary parts to form a tensor $\bm{X} \in \mathbb{R}^{L \times 2 \times M \times N}$. Each channel matrix is thus treated as a two-channel 2D image, where the channels correspond to the real and imaginary components of the signal. This representation enables the use of 2D convolutional layers while preserving the complex-valued nature of the input.

\subsection{Feature Extraction via Convolutional Downsampling}
\begin{figure}[!t]
    \centering
     \subfloat[1st TTI in TF domain]{
        		\label{fig:htf1}
    		\includegraphics[width=0.465\linewidth]{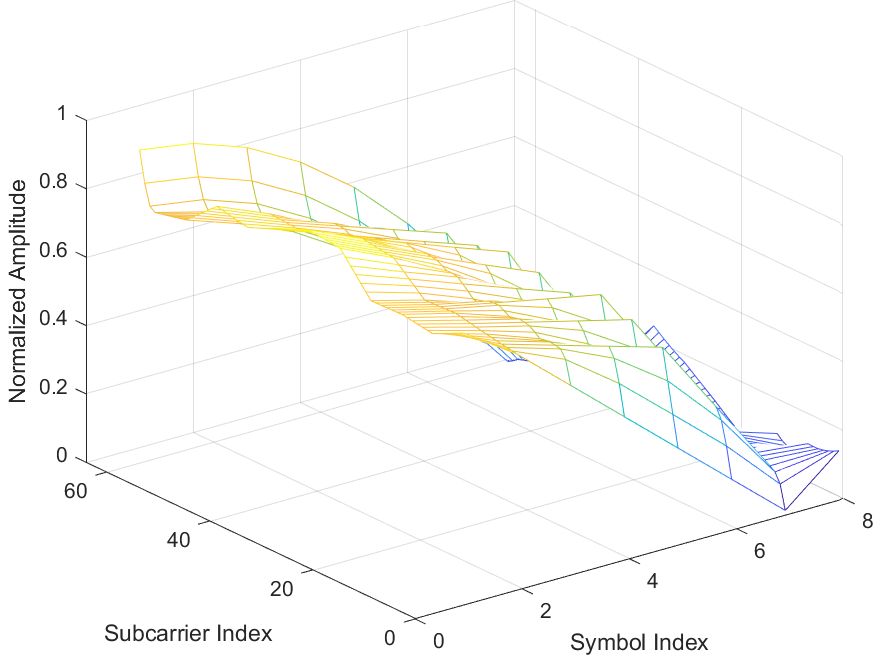}
    	}
    \hfill
         \subfloat[2nd TTI in TF domain]{
        		\label{fig:htf2}
    		\includegraphics[width=0.465\linewidth]{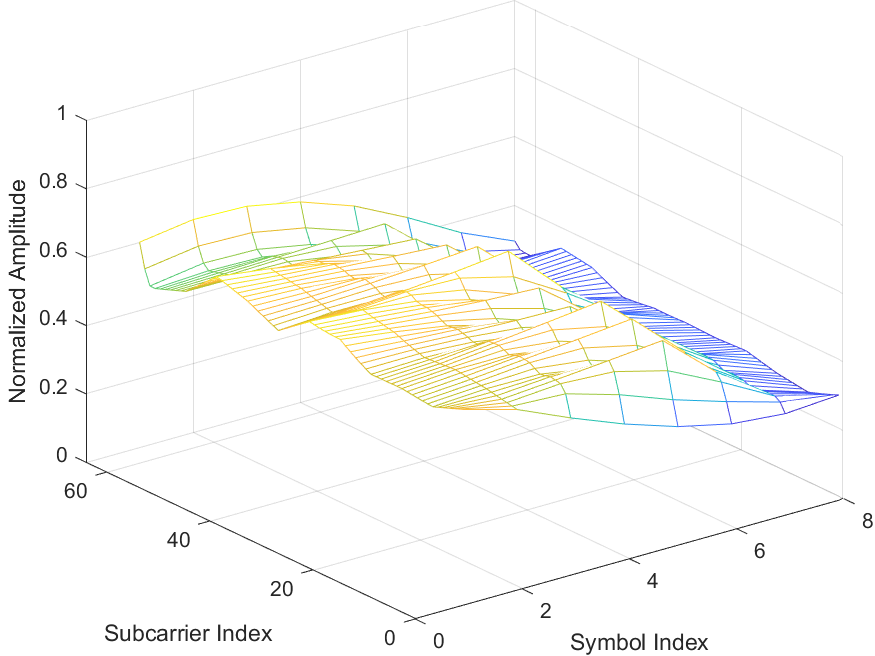}
    	}
%   \vspace{1em}

         \subfloat[1st TTI in DD domain]{
        		\label{fig:hdd1}
    	        \includegraphics[width=0.465\linewidth]{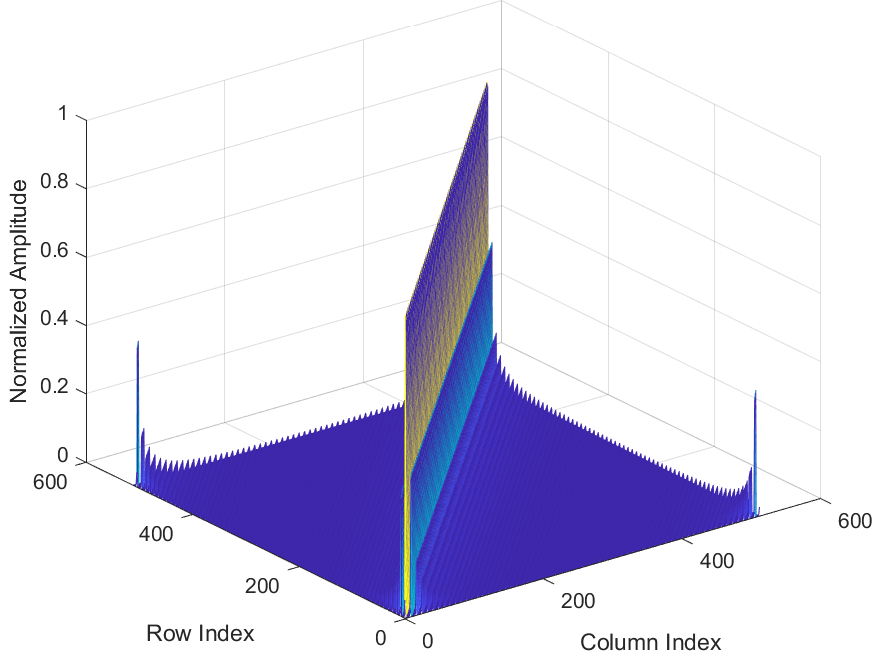}
    	}
  \hfill
           \subfloat[2nd TTI in DD domain]{
        		\label{fig:hdd2}
    	        \includegraphics[width=0.465\linewidth]{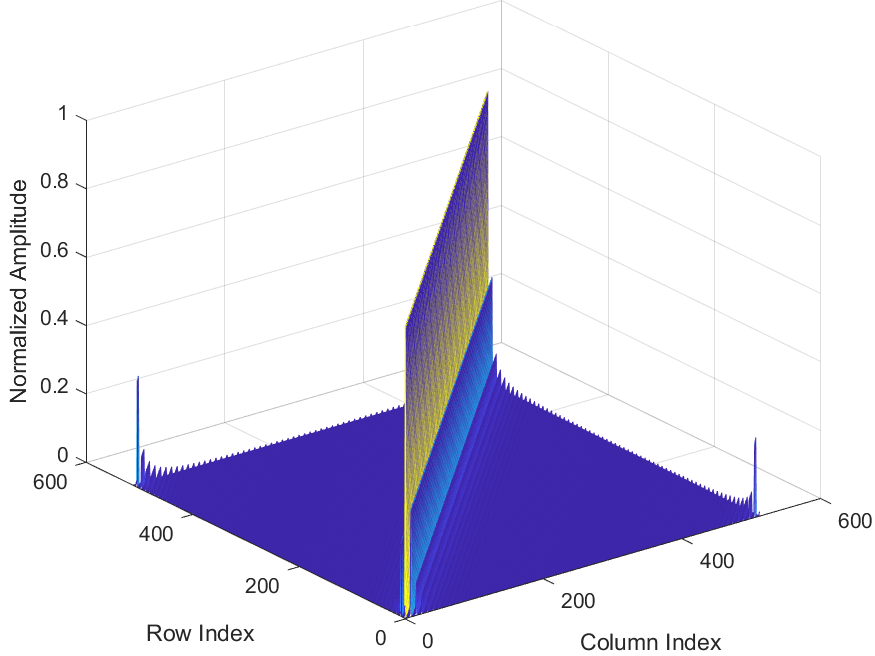}
    	}	
    \caption{Channel variation in TF and DD domains at the speed of $500$ \si{km/h}.}
    \label{fig:HTFvsHDD}
    \vspace{-5pt}
\end{figure}

The DD domain representation of OTFS channels is known to be highly sparse~\cite{OTFS,10138552}. As illustrated in Fig.~\ref{fig:HTFvsHDD}, OTFS provides enhanced channel stability under high-mobility conditions. For example, at $500$~\si{km/h}, the TF-domain channels (Fig.~\ref{fig:htf1} vs. Fig.~\ref{fig:htf2}) exhibit rapid variations between consecutive transmission time intervals (TTIs), which significantly complicates prediction. In contrast, the corresponding DD-domain channels (Fig.~\ref{fig:hdd1} vs. Fig.~\ref{fig:hdd2}) remain structurally stable and energy-concentrated, thereby simplifying prediction and improving robustness in dynamic environments.

\begin{figure}
    \centering
    \includegraphics[width=0.6\linewidth]{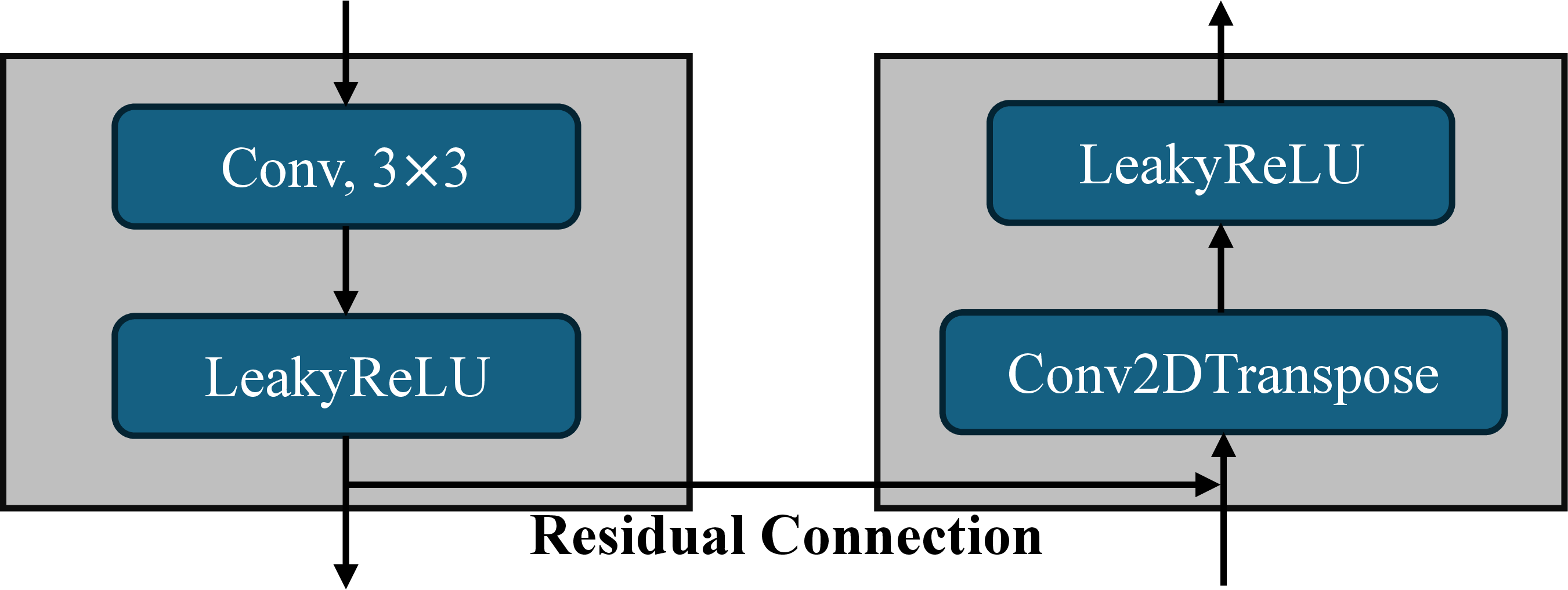}
    \vspace{-5pt}
    \caption{Downsampling convolution block and upsampling transposed convolution block.}
    \label{fig:conv_and_transpose}
    \vspace{-10pt}
\end{figure}

To exploit this sparsity, we employ convolutional downsampling blocks that reduce spatial resolution while extracting compact and informative features. Specifically, the high-dimensional $MN \times MN$ channel matrices are compressed into low-dimensional latent representations of size $R$, preserving the essential structure needed for prediction. Each downsampling block consists of a 2D convolution layer followed by a LeakyReLU activation, as illustrated in Fig.~\ref{fig:conv_and_transpose}.

\begin{figure}[!t]
    \centering
    \includegraphics[width=.5\linewidth]{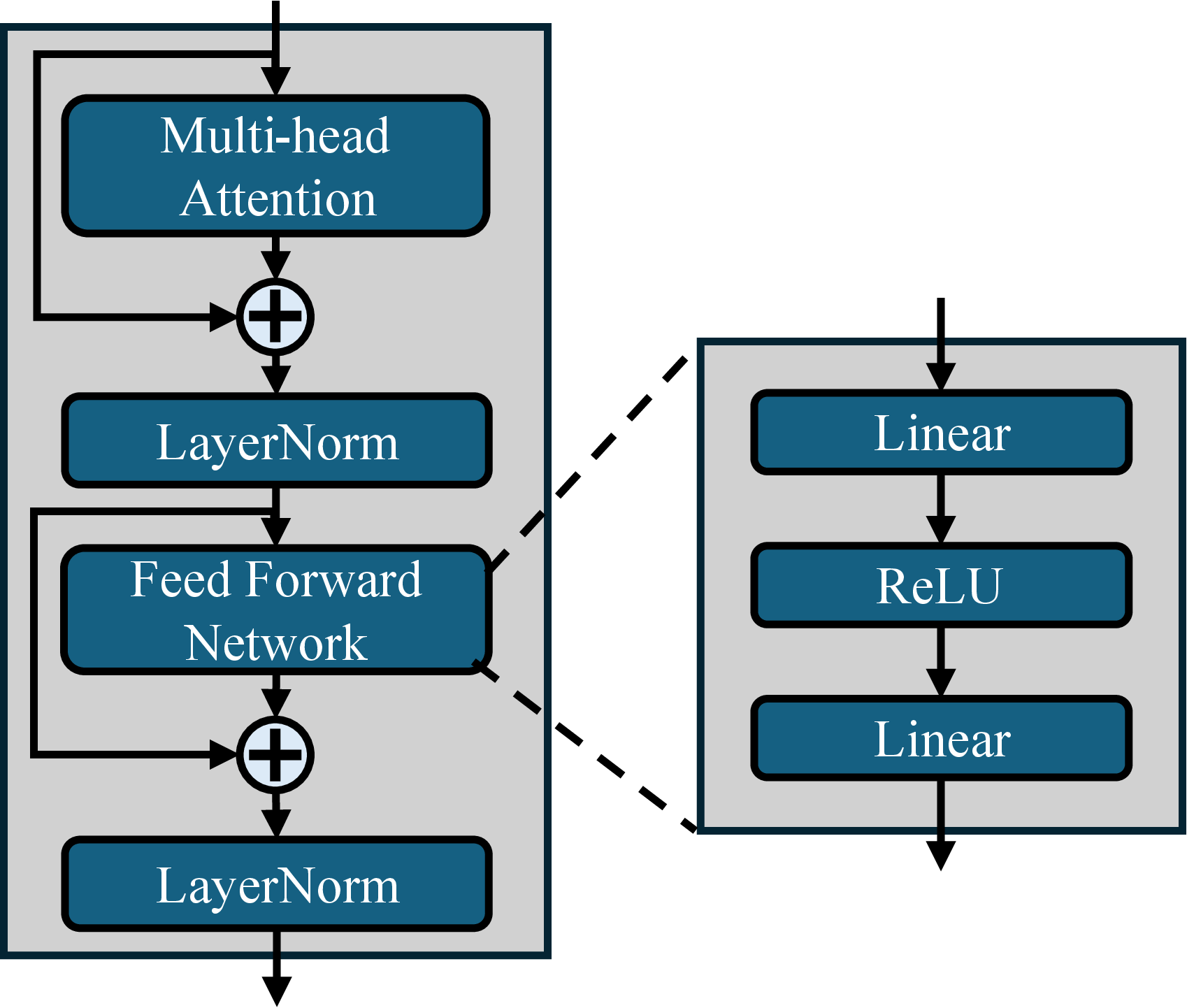}
    \vspace{-5pt}
    \caption{Temporal Transformer block structure.}
    \label{fig:blocks}
    \vspace{-10pt}
\end{figure}

\subsection{Temporal Dependency Modeling with Transformer Architecture}
After feature extraction, spatially downsampled features are flattened into temporal sequences. Learnable positional encodings preserve temporal order before feeding the sequences into stacked Transformer encoder layers (Fig.~\ref{fig:blocks}).

The Transformer’s self-attention captures complex temporal relationships across historical channel states, enabling the model to learn the dynamics of the channels. A causal mask ensures predictions depend only on current and past steps, preventing future information leakage.
Each layer comprises a multi-head self-attention mechanism and a position-wise feed-forward network, both wrapped with residual connections and layer normalization. This structure effectively models both local and global temporal dependencies in channel evolution.

\subsection{Feature Reconstruction and Channel Prediction}
The Transformer output is reshaped into 2D feature maps and processed through transposed convolutional layers to upsample from the low-dimensional representation back to the original DD-domain size, mirroring the downsampling stages. Residual connections help recover fine-grained spatial details lost during downsampling (Fig.~\ref{fig:conv_and_transpose}). The network predicts a right-shifted sequence, with the final element corresponding to the DD-domain channel at time slot $t$. It combines real and imaginary parts to reconstruct the complex-valued matrix $\hat{\bm{H}}_t$.

By learning an end-to-end mapping from historical CSI to future channel states, the model jointly optimizes compact feature extraction and temporal modeling, enabling accurate and efficient channel prediction in high-mobility OTFS systems.

In summary, a complete description of the proposed channel prediction process is formalized in Algorithm~\ref{alg:channel_prediction}. 

\begin{algorithm}[t!]
\caption{\small CNN-Transformer Channel Prediction Algorithm}
\label{alg:channel_prediction}
\small
\begin{algorithmic}[1]
\REQUIRE Historical DD channel matrices $\{\bm{H}_{t-L}, \cdots, \bm{H}_{t-1}\}$
\ENSURE Predicted channel matrix $\hat{\bm{H}}_t$
\STATE \textbf{Input Preprocessing:}
\STATE Separate real and imaginary parts: $\bm{H}_i = \text{Re}(\bm{H}_i) + j\text{Im}(\bm{H}_i)$
\STATE Form input tensor: $\bm{X} \in \mathbb{R}^{L \times 2 \times M \times N}$
\STATE \textbf{Low-Dimensional Feature Extraction:}
\FOR{each downsampling block $k = 1, 2, \cdots, K$}
    \STATE $\bm{F}_k = \text{Conv2D}(\bm{F}_{k-1})$ \COMMENT{$\bm{F}_0 = \bm{X}$}
    \STATE $\bm{F}_k = \text{LeakyReLU}(\bm{F}_k)$
\ENDFOR
\STATE Obtain compact features: $\bm{F}_{\text{low}} \in \mathbb{R}^{L \times C \times H \times W}$
\STATE \textbf{Temporal Modeling:}
\STATE Flatten spatial dimensions: $\bm{Z} = \text{Flatten}(\bm{F}_{\text{low}}) \in \mathbb{R}^{L \times D}$
\STATE Add positional encoding: $\bm{Z} = \bm{Z} + \bm{PE}$
\FOR{each Transformer layer $l = 1, 2, \cdots, L_{\text{trans}}$}
    \STATE $\bm{Z}_l = \text{MultiHeadAttention}(\bm{Z}_{l-1})$ with causal mask
    \STATE $\bm{Z}_l = \text{LayerNorm}(\bm{Z}_l + \bm{Z}_{l-1})$
    \STATE $\bm{Z}_l = \text{FFN}(\bm{Z}_l)$
    \STATE $\bm{Z}_l = \text{LayerNorm}(\bm{Z}_l + \text{input to FFN})$
\ENDFOR
\STATE \textbf{Feature Reconstruction:}
\STATE Reshape to 2D: $\bm{F}_{\text{out}} = \text{Reshape}(\bm{Z}_{L_{\text{trans}}})$
\FOR{each upsampling block $k = 1, 2, \cdots, K$}
    \STATE $\bm{G}_k = \text{ConvTranspose2D}(\bm{G}_{k-1})$ \COMMENT{$\bm{G}_0 = \bm{F}_{\text{out}}$}
    \STATE $\bm{G}_k = \text{LeakyReLU}(\bm{G}_k)$
    \STATE Apply residual connection if applicable
\ENDFOR
\STATE \textbf{Output Generation:}
\STATE Extract prediction: $\hat{\bm{H}}_t = \bm{G}_K[-1, :, :, :]$ \COMMENT{Last sequence element}
\STATE Combine channels: $\hat{\bm{H}}_t = \hat{\bm{H}}_t[:,:,0] + j\hat{\bm{H}}_t[:,:,1]$
\RETURN $\hat{\bm{H}}_t$
\end{algorithmic}
\end{algorithm}

\section{Numerical Results and Discussions}
We evaluate the proposed method using a dataset generated with the Vienna LTE-A Downlink Link Level Simulator \cite{Vienna}, under the Extended Vehicular A (EVA) channel model at a high mobility of $500$ \si{km/h} \cite{10214216}. The DD-domain channel matrices have dimensions $512 \times 512$. The model is trained using a history length of $L = 10$, batch size of $8$, and the Adam optimizer with a learning rate of $0.001$. Early stopping is employed based on validation loss to prevent overfitting. The low-dimensional latent size is set to $R = 32$, determined through sparsity analysis of the DD-domain channel. 

\begin{table}[!t]
	\centering
	\renewcommand\arraystretch{1.25}
	\caption{OTFS Channel Simulation Parameters}
	\begin{threeparttable}[!t]
	\begin{tabular}{!{\vrule width1.2pt} c !{\vrule width1.2pt} c !{\vrule width1.2pt}}
		\Xhline{1.2pt} 
		\textbf{Parameter} & \textbf{Value} \\
		\Xhline{1.2pt} 
		Speed & $500$ \si{km/h} \\
		Subcarriers & $64$ \\
		FFT Points & $64$ \\
		Symbols & $8$ \\
		Channel Model & EVA \\
		Center Frequency & $2.5$ GHz \\
		\Xhline{1.2pt} 
	\end{tabular}
	\label{tab:OTFS_channel_parameter}
	\end{threeparttable}
    \vspace{-10pt}
\end{table}

For clarity and reproducibility, the key simulation parameters are summarized in Table~\ref{tab:OTFS_channel_parameter}. We evaluate the proposed LDformer against several baseline predictors, including RepeatLast, LinearTrend, MovingAverage, TimeLinear, and DLinear~\cite{CP-DLinear}. To further assess computational efficiency, Table~\ref{tab:Model_Complexity} reports the number of trainable parameters and the average inference time per sample (for single-step prediction) of the neural-network-based models. These results show that LDformer strikes a favorable tradeoff between prediction accuracy and complexity, while remaining lightweight enough for practical deployment.

It is remarkable that, owing to the extremely high dimensionality of the input features, conventional sequential models such as LSTM become computationally infeasible. Specifically, the large number of parameters results in prohibitively high memory and runtime costs, rendering LSTM unsuitable as a baseline in this setting.

Channel prediction performance is evaluated using root mean square error (RMSE) and mean absolute error (MAE), each defined as:
\begin{align}
    \text{RMSE} &= \frac{1}{K}\sum_{i=1}^{K} \frac{1}{MN} \|\hat{\bm{H}}_{i} - \bm{H}_{i}\|_F, \\
    \text{MAE} &= \frac{1}{K}\sum_{i=1}^{K} \frac{1}{M^2N^2} \|\operatorname{vec}(\hat{\bm{H}}_{i} - \bm{H}_{i}) \|_1,
\end{align}
where $\hat{\bm{H}}_{i}$ and $\bm{H}_{i}$ denote the predicted and true channel matrices of size $MN \times MN$, respectively, and $K$ represents the number of matrix samples.

\begin{table}[!t]
	\centering
 	\renewcommand\arraystretch{1.25}
	\caption{Model Parameters and Sample Inference Time}
    	\begin{threeparttable}[!t]
	\begin{tabular}{!{\vrule width1.2pt} c !{\vrule width1.2pt} c !{\vrule width1.2pt} c !{\vrule width1.2pt}}
		\Xhline{1.2pt} 
		\textbf{Model} & \textbf{Parameters (M)} & \multirow{1}{*}{\textbf{Inference Time (ms)}} \\
		\Xhline{1.2pt} 
		Proposed & 25.62 & 10.63 \\
		\hline
		DLinear & 11.54 & 1.58 \\
		\hline 
		TimeLinear (MLP) & 5.77 & 0.43 \\
		\Xhline{1.2pt}
	\end{tabular}
	\label{tab:Model_Complexity}
    	\end{threeparttable}
    \vspace{-10pt}
\end{table}

\begin{figure}[!t]
    \centering
    \includegraphics[width=0.85\linewidth]{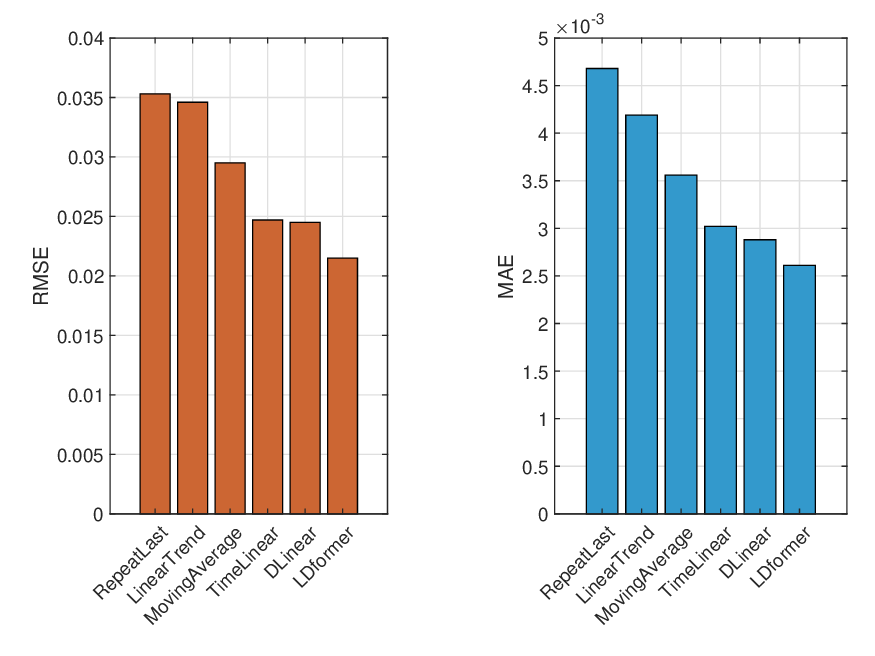}
    \vspace{-10pt}
    \caption{Performance comparison between LDformer and baseline models.}
    \label{fig:OTFS_perf}
    \vspace{-10pt}
\end{figure}

The comparison results are presented in Fig.~\ref{fig:OTFS_perf}. Our LDformer model consistently outperforms all baseline methods, achieving the lowest RMSE of $0.0215$ and MAE of $0.00261$. Compared to the second-best baseline, DLinear, LDformer reduces RMSE by $12.2\%$ and MAE by $9.4\%$. These gains are particularly notable given the extremely high-mobility conditions, where rapid channel variations significantly complicate prediction.

\begin{figure}[!t]
	\centering
	\includegraphics[width=.7\linewidth]{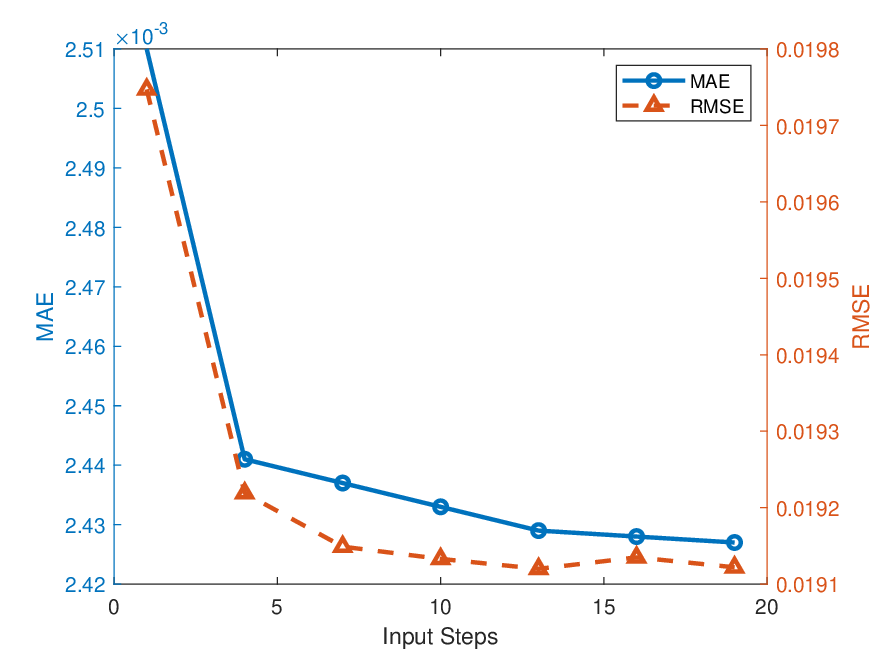}
    \vspace{-10pt}
	\caption{Effect of input history length $L$ on prediction errors (MAE/RMSE).}
	\label{fig:diff_L}
    \vspace{-10pt}
\end{figure}

The proposed model supports flexible adjustment of both input history length and prediction horizon during inference, enabling adaptation to diverse forecasting requirements. We evaluate robustness under varying input lengths $L$ and multi-step prediction horizons. Trained with single-step prediction at $L=20$, LDformer benefits from longer input histories at small-to-moderate values, while the performance gain saturates as $L$ increases (Fig.~\ref{fig:diff_L}). 

\begin{figure}[!t]
	\centering
		\subfloat[Prediction errors vs. prediction horizon]{\label{fig:multi_step_error}\includegraphics[width=0.49\linewidth]{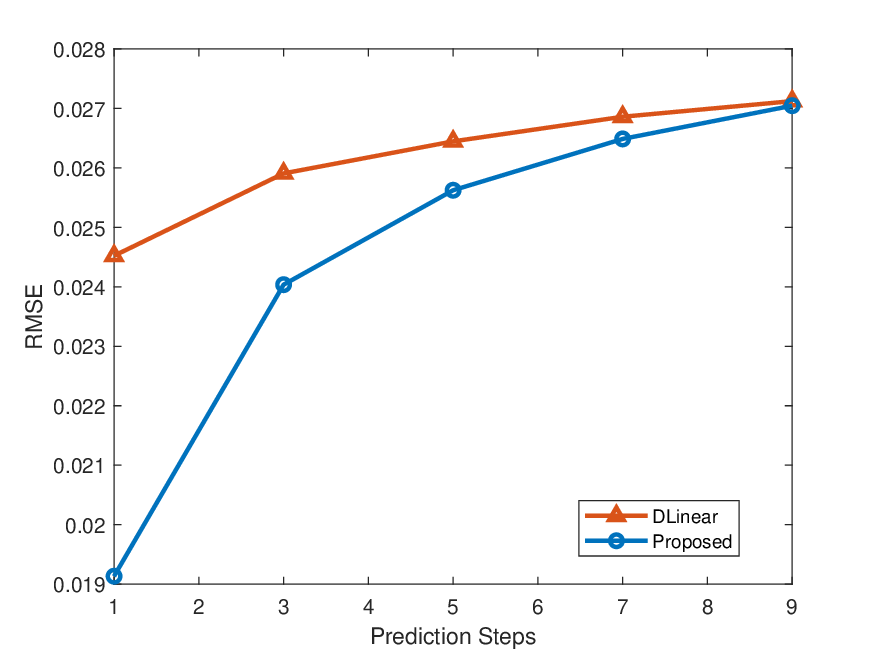}}
		\hfill
		\subfloat[Inference time vs. prediction horizon]{\label{fig:multi_step_inference_time}\includegraphics[width=0.49\linewidth]{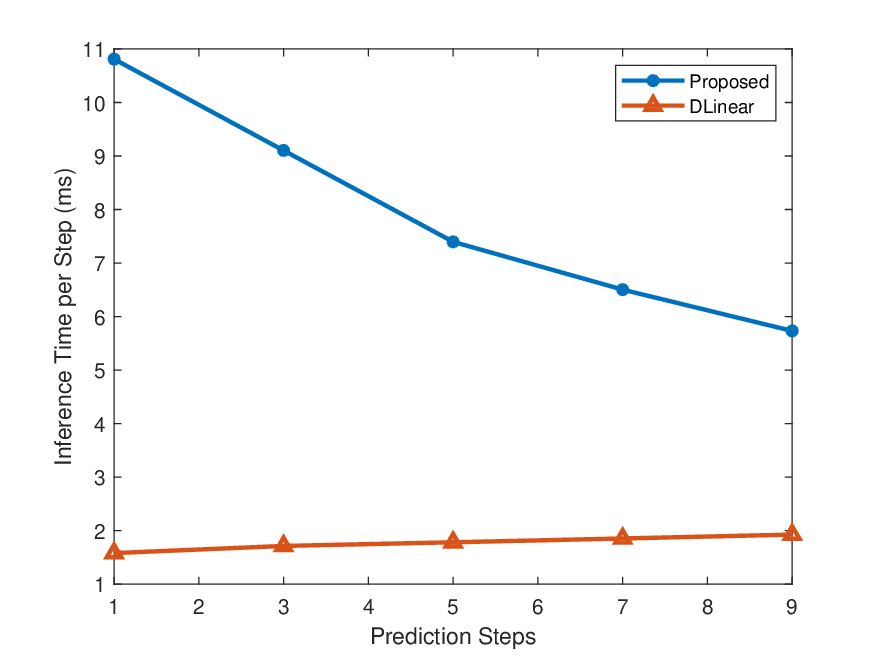}}
		\caption{Multi-step prediction comparison for LDformer and DLinear. The inference time is measured on the entire test set, which contains around $2000$ samples.}
    \vspace{-10pt}
\end{figure}

For multi-step prediction, LDformer consistently outperforms DLinear in terms of error. Although its per-step inference time is slightly higher, the internal autoregressive mechanism ensures that the average inference time scales sub-linearly with the prediction horizon, thereby maintaining efficiency for long-horizon forecasting (Figs.~\ref{fig:multi_step_error} and \ref{fig:multi_step_inference_time}).

\begin{figure}[!t]
	\centering
	\includegraphics[width=.7\linewidth]{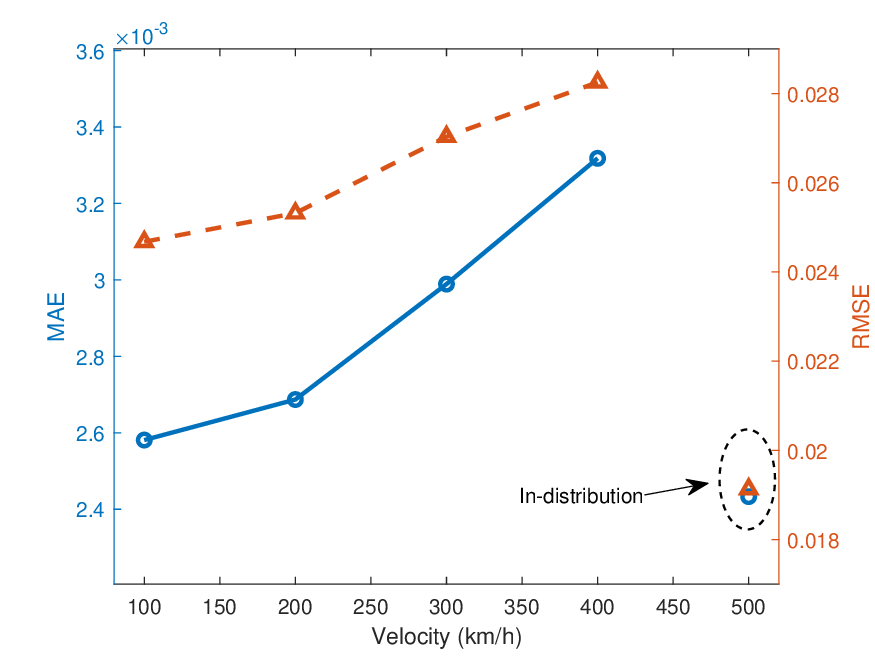}
    \vspace{-10pt}
	\caption{Cross-speed generalization performance of the trained model.}
	\label{fig:cross_speed}
    \vspace{-10pt}
\end{figure}

Cross-speed generalization is evaluated by training at a speed of $500$~\si{km/h} and testing on speeds ranging from $100$ to $500$~\si{km/h}. While LDformer maintains strong in-distribution performance, prediction accuracy degrades at unseen lower speeds, with both MAE and RMSE increasing approximately monotonically with speed mismatch (Fig.~\ref{fig:cross_speed}). These results indicate that strategies such as mixed-speed training, lightweight fine-tuning, or explicit speed conditioning could further enhance robustness across diverse mobility regimes.

\section{Conclusions}
This paper proposed a channel prediction framework for OTFS-based high-mobility wireless systems, addressing the strict reliability and latency requirements of URLLC-enabled IoT applications. By exploiting DD-domain sparsity and modeling temporal dynamics with a hybrid CNN-Transformer architecture, LDformer achieves high prediction accuracy while maintaining computational efficiency. Simulations under extreme $500$ \si{km/h} mobility demonstrate its effectiveness, with RMSE and MAE reduced by $12.2\%$ and $9.4\%$, respectively, compared to the competitive DLinear baseline. These results highlight the advantages of DD-domain representations in fast-varying environments. Future work includes real-time deployment, model compression for edge devices, and extensions to multi-antenna OTFS systems.

\bibliographystyle{IEEEtran}
\balance
\bibliography{Reference}
\end{document}